\documentclass{webofc}
\usepackage[varg]{txfonts}   
\usepackage{graphicx,amsmath,amssymb}
\usepackage{caption,subcaption}
\usepackage{xcolor}
\usepackage{fancyhdr}

\graphicspath{{./img/}}

\begin{document}
\title{Minimal 4D Composite Higgs Models Under Current LHC Constraints}
%
%

\author{
\firstname{Ethan} 
\lastname{Carragher}
\inst{1}\fnsep\thanks{Speaker, \email{ethan.carragher@adelaide.edu.au}} \and
\firstname{Daniel} \lastname{Murnane}\inst{1}\fnsep\thanks{\email{daniel.murnane@adelaide.edu.au}} \and
\firstname{Peter} \lastname{Stangl}\inst{2} \fnsep\thanks{\email{peter.stangl@lapth.cnrs.fr}}
\and
\firstname{Wei} \lastname{Su}\inst{1}
\fnsep\thanks{\email{wei.su@adelaide.edu.au}} \and
\firstname{Martin} \lastname{White}\inst{1}
\fnsep\thanks{\email{martin.white@adelaide.edu.au}} \and
\firstname{Anthony G.} \lastname{Williams}\inst{1}
\fnsep\thanks{\email{anthony.williams@adelaide.edu.au}}
}
\institute{ARC Centre of Excellence for Particle Physics at the Terascale, Department of Physics,\\University of Adelaide, South Australia 5005, Australia 
\and
    Laboratoire d'Annecy-le-Vieux de Physique Th{\'e}orique, UMR5108, CNRS, F-74941, Annecy-le-Vieux Cedex, France 
    }

\abstract{%
  We present preliminary results of the first convergent global fits of several minimal composite Higgs models. Our fits are performed using the differential evolution optimisation package \texttt{Diver}. A variety of physical constraints are taken into account, including a wide range of exclusion bounds on heavy resonance production from Run 2 of the LHC. As a by-product of the fits, we analyse the collider phenomenology of the lightest new up-type and down-type resonances in the viable regions of our models, finding some low-mass resonances that can be probed in future collider searches.
}

\maketitle
\thispagestyle{fancy}
\section{Introduction}
\label{Introduction}
Theories that realise the Higgs boson as a bound state of some new strong dynamics, rather than as an elementary particle, are attractive solutions to the Higgs mass hierarchy problem. In such composite Higgs models (CHMs), the Higgs emerges as a pseudo-Nambu-Goldstone boson of some spontaneously broken symmetry so that it is naturally lighter than other bound states of the same strong dynamics. This so-called composite sector is expected to be at the few-TeV scale, leading to the exciting prospect that the LHC will soon be able to find evidence for, or rule out, certain CHMs. 

We focus in this work on minimal CHMs, based on the $S\! O(5) \rightarrow S\! O(4)$ symmetry breaking pattern \cite{agashe2005}. Our goal is to perform the first convergent global fits of three different minimal models (specified in Sect.~\ref{Models}), finding the viable regions of the models' parameter spaces given a wide range of physical constraints. Although the subject of much theoretical work~\cite{de2012,panico2011,panico2012,marzocca2012,matsedonskyi2012,carmona2015}, such models have so far resisted global fits on account of their large parameter spaces and highly non-trivial parameter dependencies. Similar models have previously been numerically explored \cite{carena2014,Niehoff:2015iaa,Barnard:2015ryq}, though when only Run 1 results from the LHC were available. We build on the strategy of Ref.~\cite{Niehoff:2015iaa} using updated constraints, including an additional 40 LHC searches at $\sqrt{s}=13$~TeV that place significantly stronger bounds on the production of heavy resonances.

To facilitate our fits we use a differential evolution optimisation algorithm that proves particularly effective at maximising difficult likelihood functions over high-dimensional spaces \cite{storn1997differential}. From these fits, experimental signatures in the viable regions of each model can be analysed, and we focus in these proceedings on the signals of the lightest up-type and down-type composite resonances.

\section{Models}
\label{Models}
\rhead{}

Minimal CHMs have some freedom in the exact structures of their composite sectors. For our work we consider models with a two-site structure that is sufficient for a finite and calculable Higgs potential, generally known as Minimal 4D CHMs (M4DCHMs) \cite{de2012,marzocca2012}. A particular M4DCHM is completely determined by specification of the $S\! O(5)$ representations under which its composite fermions transform. In the interests of minimising the parameter spaces of our models, we take the limiting case where out of the SM fermions, only the third generation quarks couple to the composite sector, for the lighter fermions are expected to couple only weakly. Accordingly, we label our models as M4DCHM$^{\mathbf{q}-\mathbf{t}-\mathbf{b}}$, where $\mathbf{q}, \mathbf{t}, \mathbf{b}$ are the $S\! O(5)$ representations under which the composite partners of the elementary $q^{0}_{L}$, $t^{0}_{R}$, $b^{0}_{R}$ respectively transform. Focusing only on representations that provide custodial protection for the $Z \bar{b}_L b_{L}$ coupling, we consider in particular the M4DCHM$^{5-5-5}$, the M4DCHM$^{14-14-10}$, and the M4DCHM$^{14-1-10}$, all of which have been detailed in Ref.~\cite{carena2014}.

In these models the Higgs field interacts exclusively through the quantity $\sin(h/f)$, where $f$ is the Higgs boson decay constant. One of the more consequential parameters, $f$ can also be regarded as the scale of symmetry breaking and defines the cutoff $\Lambda = 4 \pi f$ of the theory. We scan over the range $f \in [0.5 \text{ TeV}, 5.0 \text{ TeV}]$. Among the remaining model parameters are three other decay constants, composite gauge couplings, elementary-composite mixing strengths, on-diagonal and off-diagonal composite masses, and Yukawa-like couplings. In total, our models have dimensionalities of 19, 17, and 15, in the order given.

\section{Fit Procedure}
\label{Procedure}

Finding the parameter values that best fit observation is equivalent to finding those points $\mathbf{p}$ that maximise some likelihood function $L$, which we take as being a multivariate Gaussian function in the observables:
\begin{align}
    L(\mathbf{p}) = e^{- \frac{1}{2} \chi^2(\mathbf{p})}, \qquad \chi^2(\mathbf{p}) = ( \mathbf{\mathcal{O}}^{\text{theo}}(\mathbf{p}) - \mathbf{\mathcal{O}}^{\text{exp}} )^\intercal C^{-1} ( \mathbf{\mathcal{O}}^{\text{theo}}(\mathbf{p}) - \mathbf{\mathcal{O}}^{\text{exp}} ).
\label{eq:likelihood_definition}
\end{align}
Here, $C$ is the covariance matrix taking into account all uncertainties, and also correlations between observables. Most observables $\mathcal{O}_i$ are not correlated with any other, having the simple additive contribution
\begin{align}
    \chi^2_i (\mathbf{p}) = \frac{\left(\mathcal{O}^{\text{theo}}_i (\mathbf{p}) - \mathcal{O}^{\text{exp}}_i \right)^2}{\sigma^2_i},
\label{eq:chi2_contribution}
\end{align}
where $\sigma_i$ is the total uncertainty of the observable. Our treatment of the constraints builds on that of Ref.~\cite{Niehoff:2015iaa} (and the subsequent modifications of Ref.~\cite{niehoff2017electroweak}), where we employ only those constraints that are applicable to a third-quark-generation-only model; namely, the Standard Model masses, the electroweak scale $v \approx 246$~GeV, the Peskin-Takeuchi S and T parameters, Z boson decay ratios, Higgs signal strengths, and direct collider searches for new resonances. For the collider searches, $\mathcal{O}^{\text{exp}}_i$ is taken as the Gaussian central value of the observed and expected upper bounds for the cross section of the given process. In the case where the observed bound is stronger than the expected bound, the contribution is modified to be
\begin{align}
    \chi^2_i (\mathbf{p}) = \frac{\left(\mathcal{O}^{\text{theo}}_i (\mathbf{p}) - \mathcal{O}^{\text{exp}}_i \right)^2 - \left(\mathcal{O}^{\text{exp}}_i \right)^2}{\sigma^2_i},
\label{eq:chi2_contribution}
\end{align}
so that a vanishing cross section gives the highest likelihood.

We use differential evolution to maximise $L$, from an initial ``population" of points $(\mathbf{p}_{i})_{i=1}^{N}$ ``breeding" successive ``generations" of points that migrate towards better-fit regions in a manner analogous to natural selection. Specifically, we use the $\lambda$jDE prescription provided by the differential evolution package \texttt{Diver} \cite{Workgroup:2017htr}. We use quite large populations of $N=50,000$ points, and declare our scans converged when the average fractional improvement in the log likelihood over the last ten generations falls below the rather strong threshold of $10^{-5}$.

\section{Global Fit Results}
\label{Results}

\begin{figure}[h]
\centering
\begin{subfigure}{.49\linewidth}
  \centering
  \includegraphics[width=1\textwidth, clip]{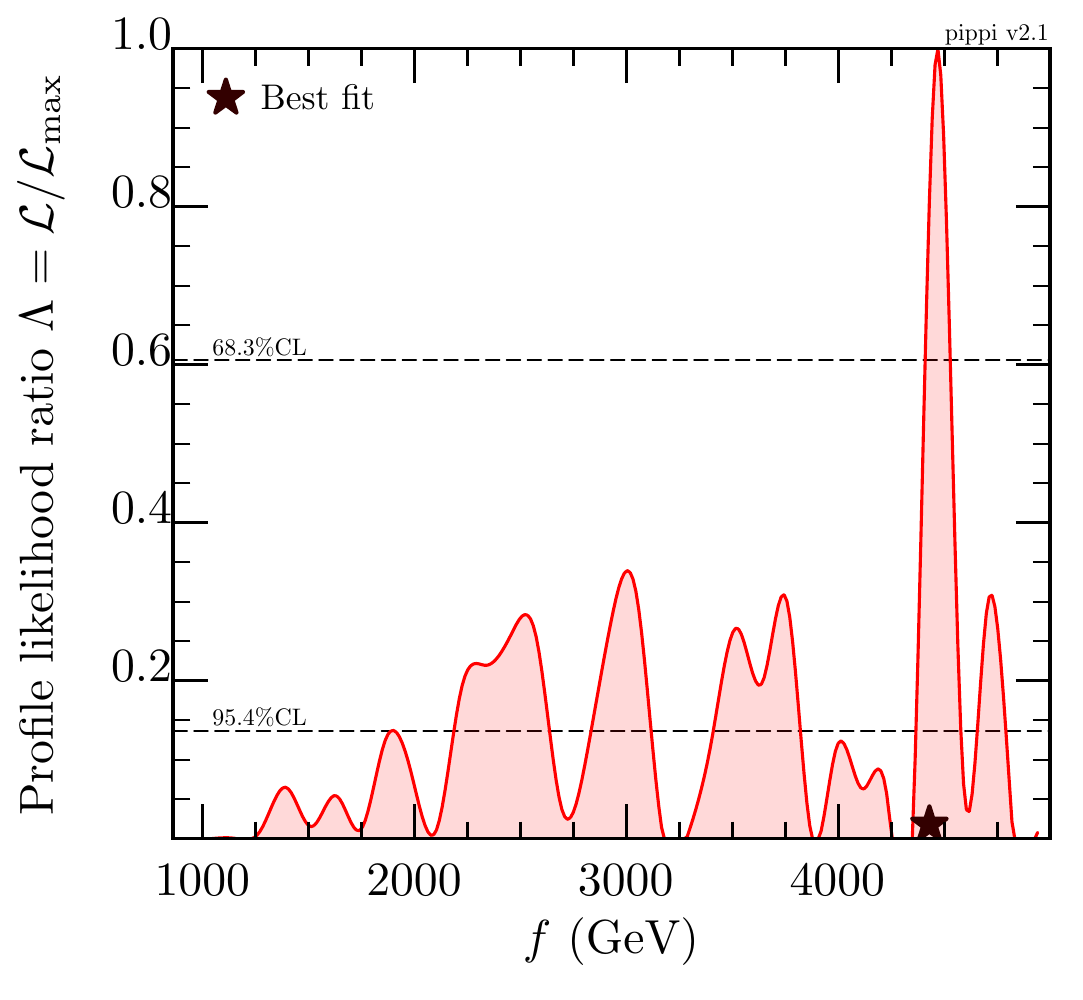}
   \caption{}
\end{subfigure}
\begin{subfigure}{.49\linewidth}
  \centering
  \includegraphics[width=1\textwidth, clip]{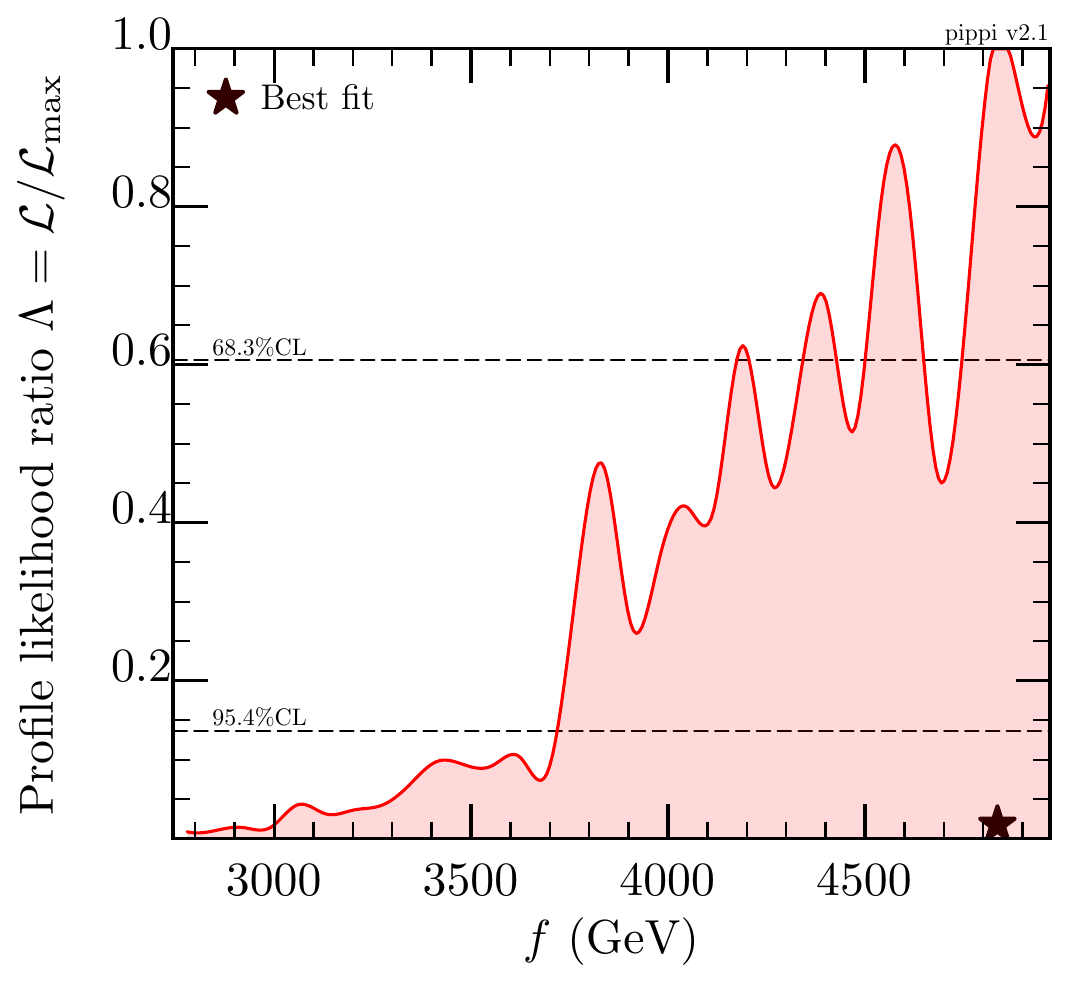}
   \caption{}
\end{subfigure}
\begin{subfigure}{.49\linewidth}
  \centering
  \includegraphics[width=1\textwidth, clip]{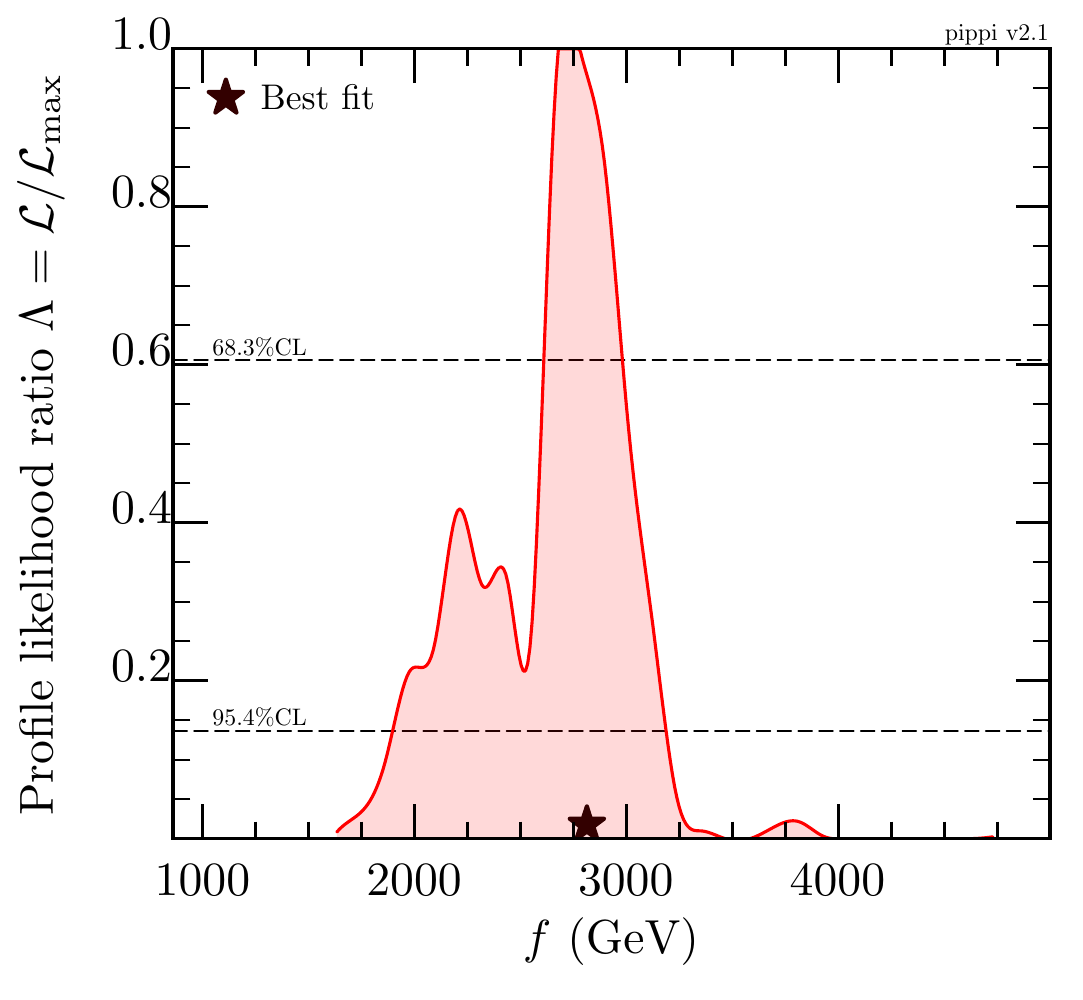}
   \caption{}
\end{subfigure}
\setlength{\belowcaptionskip}{0pt}
\caption{Profile likelihood ratios of the Higgs decay constant $f$ in (a) the M4DCHM$^{5-5-5}$, (b) the M4DCHM$^{14-14-10}$, (c) the M4DCHM$^{14-1-10}$. For reference, the best-fit points found in these models have respective negative log likelihoods of $15.3$, $15.1$, and $15.4$.}
\label{fig:f_likelihood_profiles}
\end{figure}

Results of the global fits are presented as profile likelihood ratios (PLRs), which express the maximum likelihood at a given parameter value in units of the global maximum likelihood. PLRs for the Higgs boson decay constant, $f$, are shown in Fig.~\ref{fig:f_likelihood_profiles}. Confidence intervals for $f$ are simply taken as those values for which the PLR lies above the significance-level dependent values marked in Fig.~\ref{fig:f_likelihood_profiles}. Note that this is the correct procedure for finding the confidence intervals for the \textit{data}, as a consequence of Wilk's theorem, since the likelihood function is Gaussian in the data, but for input parameters such as $f$ this prescribes confidence intervals that are only approximate, due to the non-linear transformations relating the data to the input parameters \cite{Akrami:2010cz,Strege:2012kv}. We have not performed coverage tests to determine the accuracy of the confidence intervals quoted for $f$ on account of the computing expense required.

Ideally, the PLRs should be smoothly varying to signal the likelihood function has been well-explored. This is approximately the case for the M4DCHM$^{14-14-10}$ and the M4DCHM$^{14-1-10}$, but the sporadic peaks in the PLR for the M4DCHM$^{5-5-5}$ indicate this model has been poorly sampled. The difficulty in fitting this model is not entirely surprising on account of its notorious ``double-tuning", which is not present in the other models \cite{panico2012,matsedonskyi2012}.

We see qualitatively similar distributions for $f$ in the M4DCHM$^{5-5-5}$ and the M4DCHM$^{14-14-10}$, with larger values of $f$ tending to result in greater likelihoods, having values $f \approx 4.5$~TeV and $f \approx 4.8$~TeV at the respective best-fit points. The non-linearities from the pseudo-Nambu-Goldstone boson nature of the Higgs evidently must be highly suppressed to best fit the data. The plots show that at the $2\sigma$ confidence level, $f \gtrsim 2.1$~TeV and $f \gtrsim 3.7$~TeV in the respective models\footnote{It seems likely that had we extended our bound on $f$ to greater than $5$~TeV, we would have found points that even better fit the data. However, as long as the best-fit points in our scans have comparable likelihoods to the true global optima, these confidence intervals will be approximately accurate.}. Note, however, that the fine-tuning typically scales as $\sim f^{2} / v^{2}$, so the models are less attractive as solutions to the hierarchy problem as $f$ is pushed to higher values. This is not too much of an issue in the M4DCHM$^{14-1-10}$, which is seen to have $f$ localised between $\sim 1.9$~TeV and $\sim 3.2$~TeV at the $2\sigma$ confidence level.

\begin{figure}[h]
\centering
\begin{subfigure}{.49\linewidth}
  \centering
  \includegraphics[width=1\textwidth, clip]{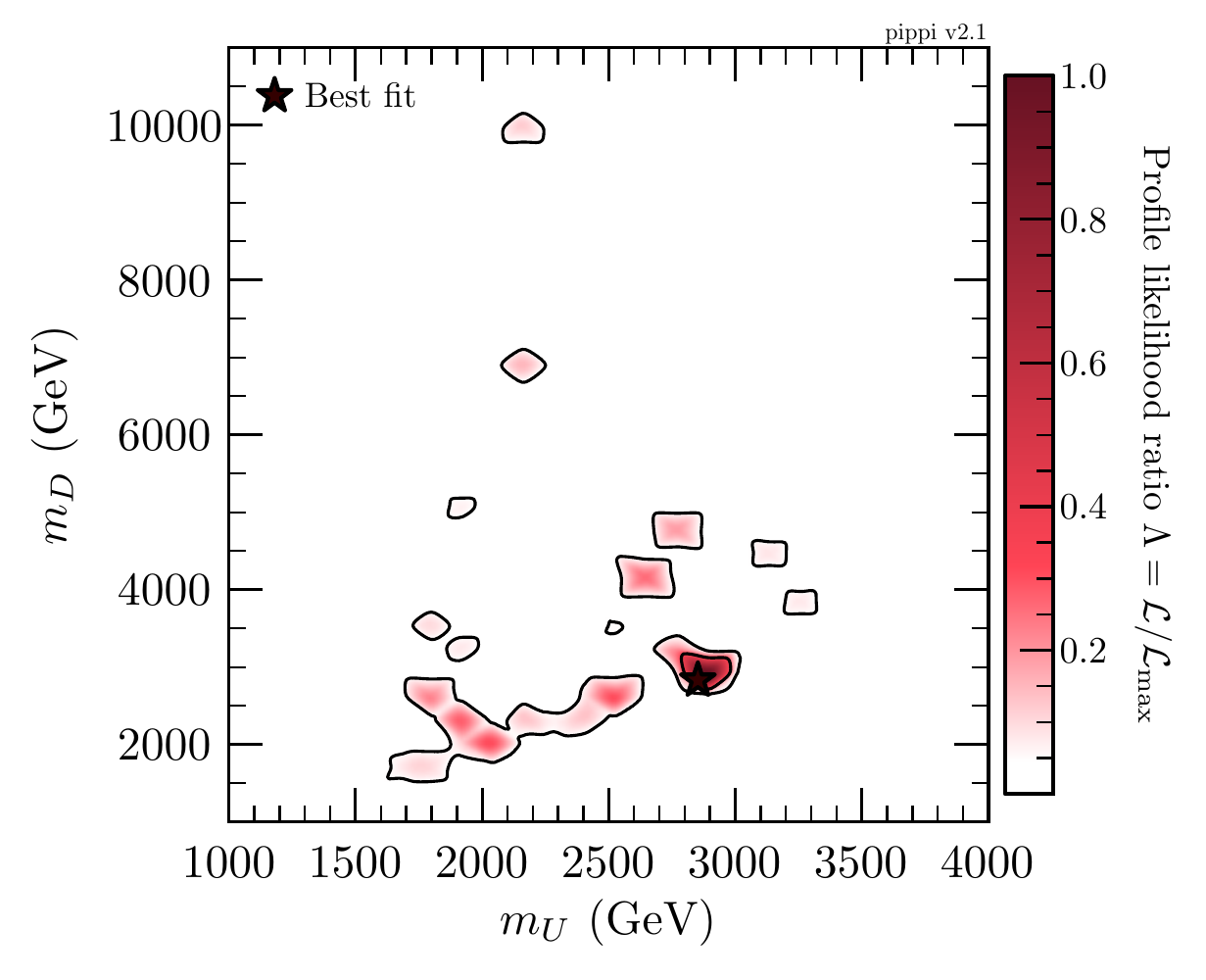}
   \caption{}
\end{subfigure}
\begin{subfigure}{.49\linewidth}
  \centering
  \includegraphics[width=1\textwidth, clip]{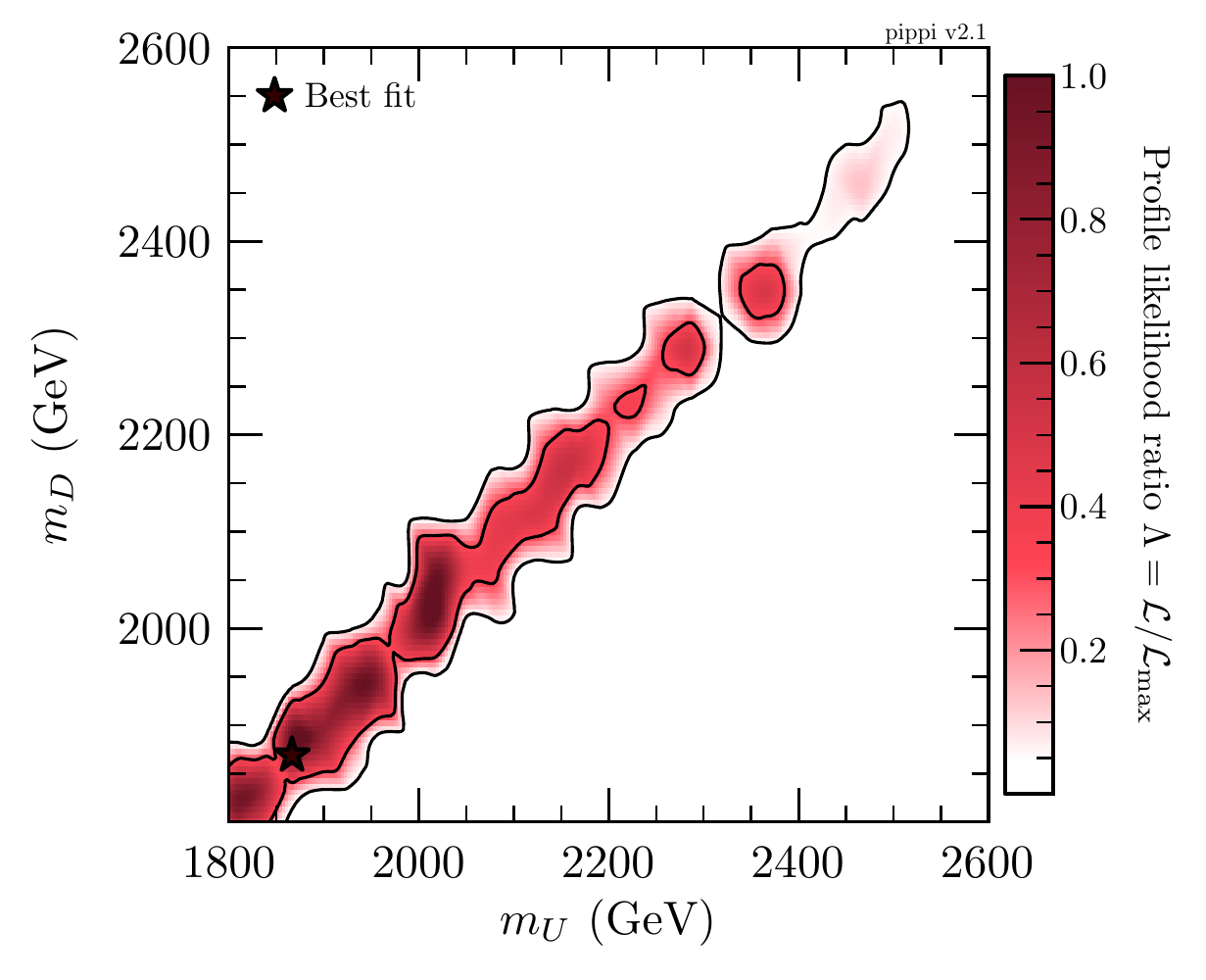}
   \caption{}
\end{subfigure}
\begin{subfigure}{.49\linewidth}
  \centering
  \includegraphics[width=1\textwidth, clip]{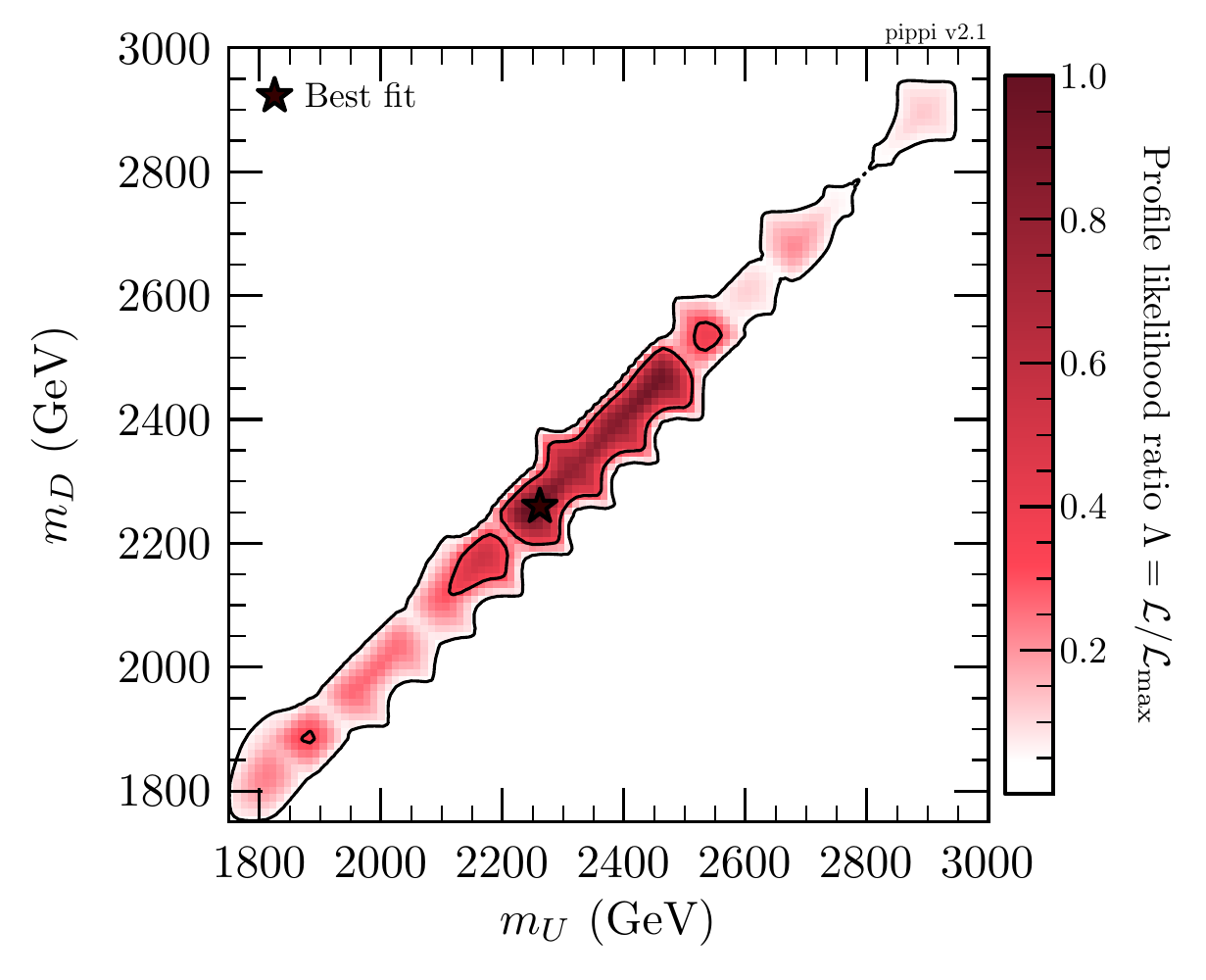}
   \caption{}
\end{subfigure}
\caption{Profile likelihood ratios of the masses of the lightest up-type (U) and down-type (D) composite resonances in (a) the M4DCHM$^{5-5-5}$, (b) the M4DCHM$^{14-14-10}$, (c) the M4DCHM$^{14-1-10}$.}
\label{fig:mass_likelihood_profiles}
\end{figure}

Next we analyse the expected collider phenomenology in the most likely regions of our models - specifically that of the lightest up-type (U) and down-type (D) composite fermionic resonances. Fig.~\ref{fig:mass_likelihood_profiles} shows the profile likelihood ratios of their masses, which are seen to be approximately degenerate in the minimally-tuned models, ranging from $1.8$~TeV to $2.5$~TeV in the M4DCHM$^{14-14-10}$, and $1.8$~TeV to $3.0$~TeV in the M4DCHM$^{14-1-10}$ at the $2\sigma$ level. These correlations are to be expected by the symmetries of the models. The M4DCHM$^{5-5-5}$, on the other hand, has a region with roughly degenerate U and D resonances, but other regions where the D resonance is by far the heavier of the two.

\begin{figure}[h]
\centering
\begin{subfigure}{.49\linewidth}
  \centering
  \includegraphics[width=1\textwidth, clip]{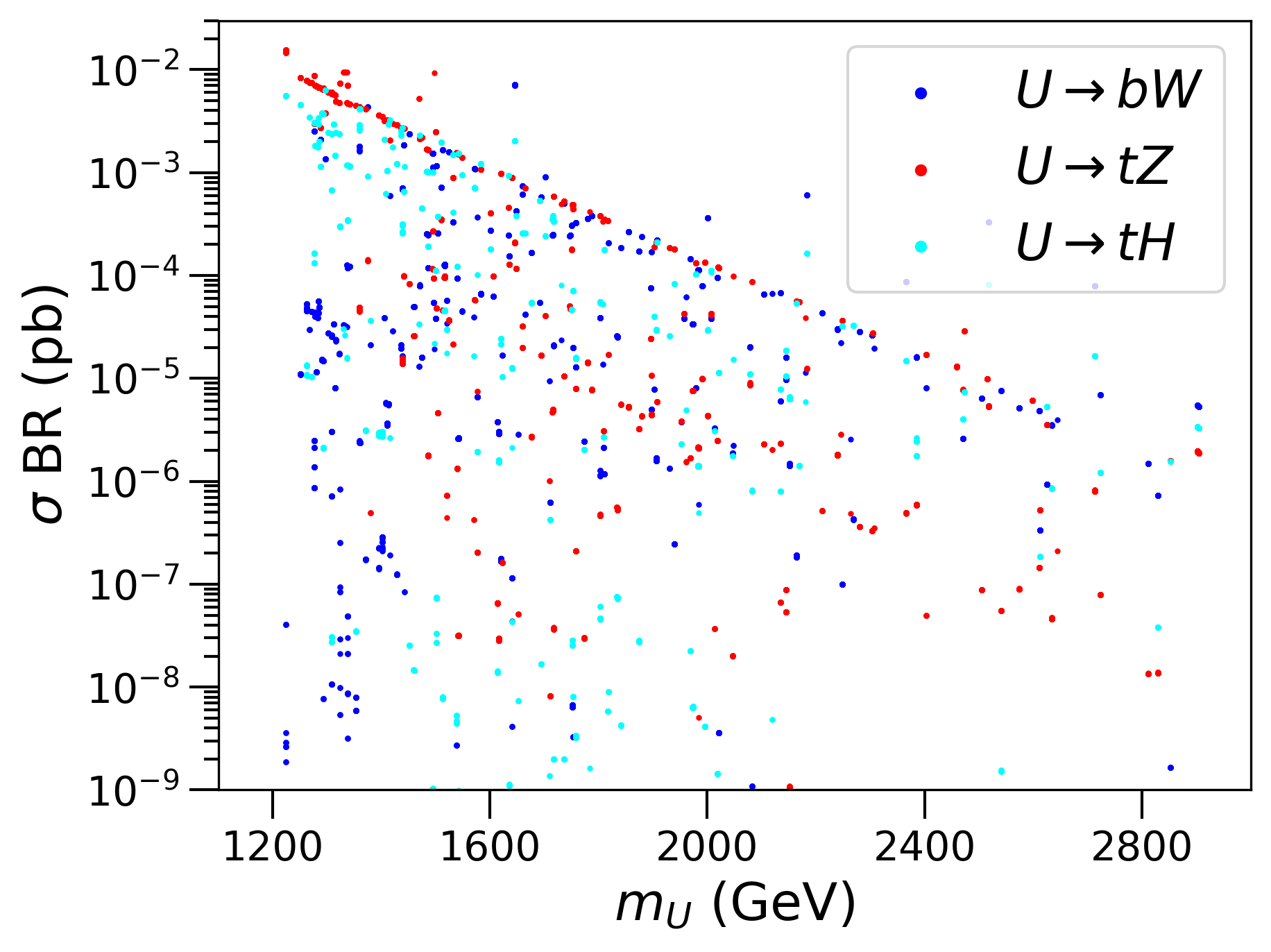}
\end{subfigure}
\begin{subfigure}{.49\linewidth}
  \centering
  \includegraphics[width=1\textwidth, clip]{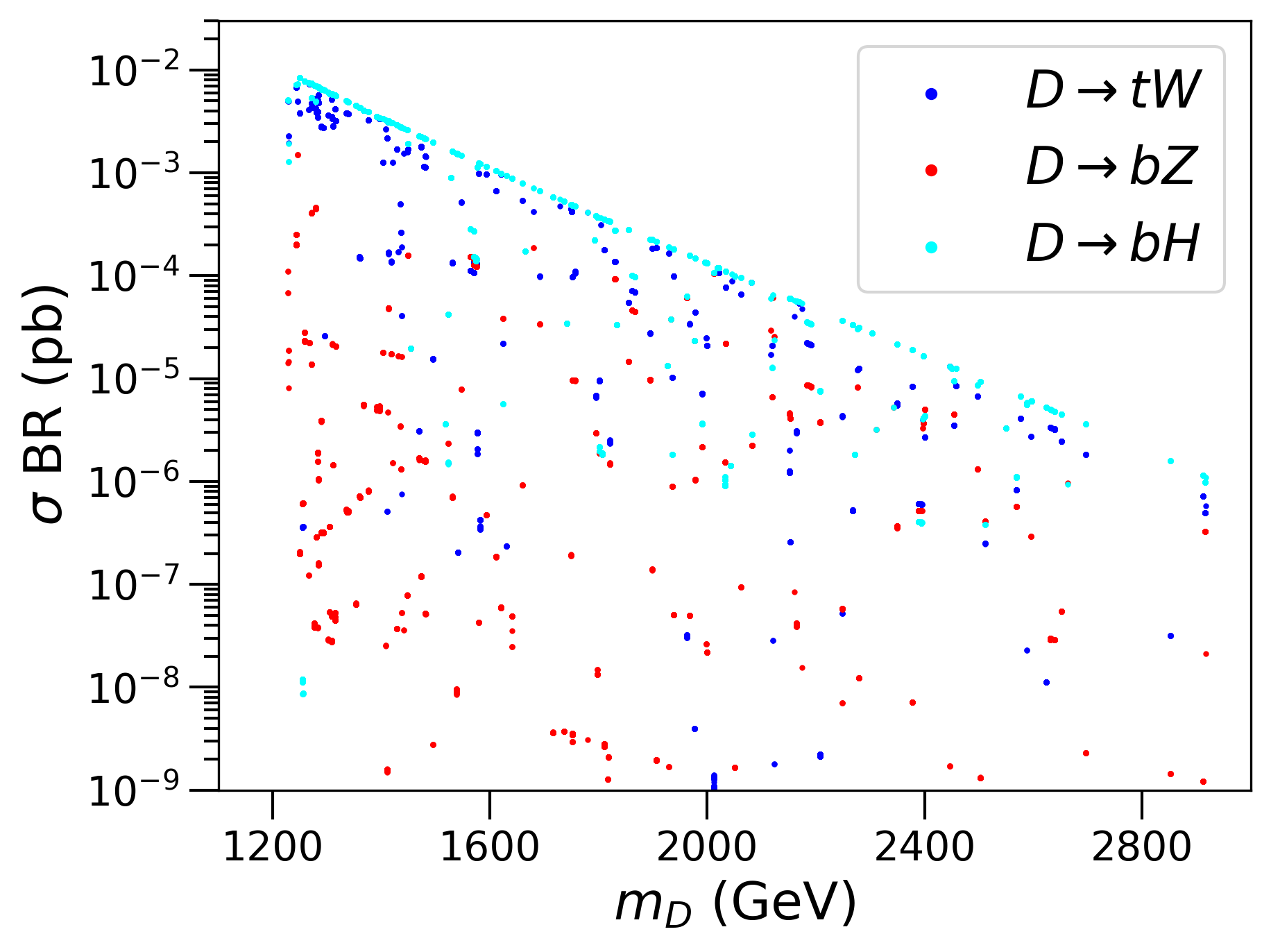}
\end{subfigure}
\begin{subfigure}{.49\linewidth}
  \centering
  \includegraphics[width=1\textwidth, clip]{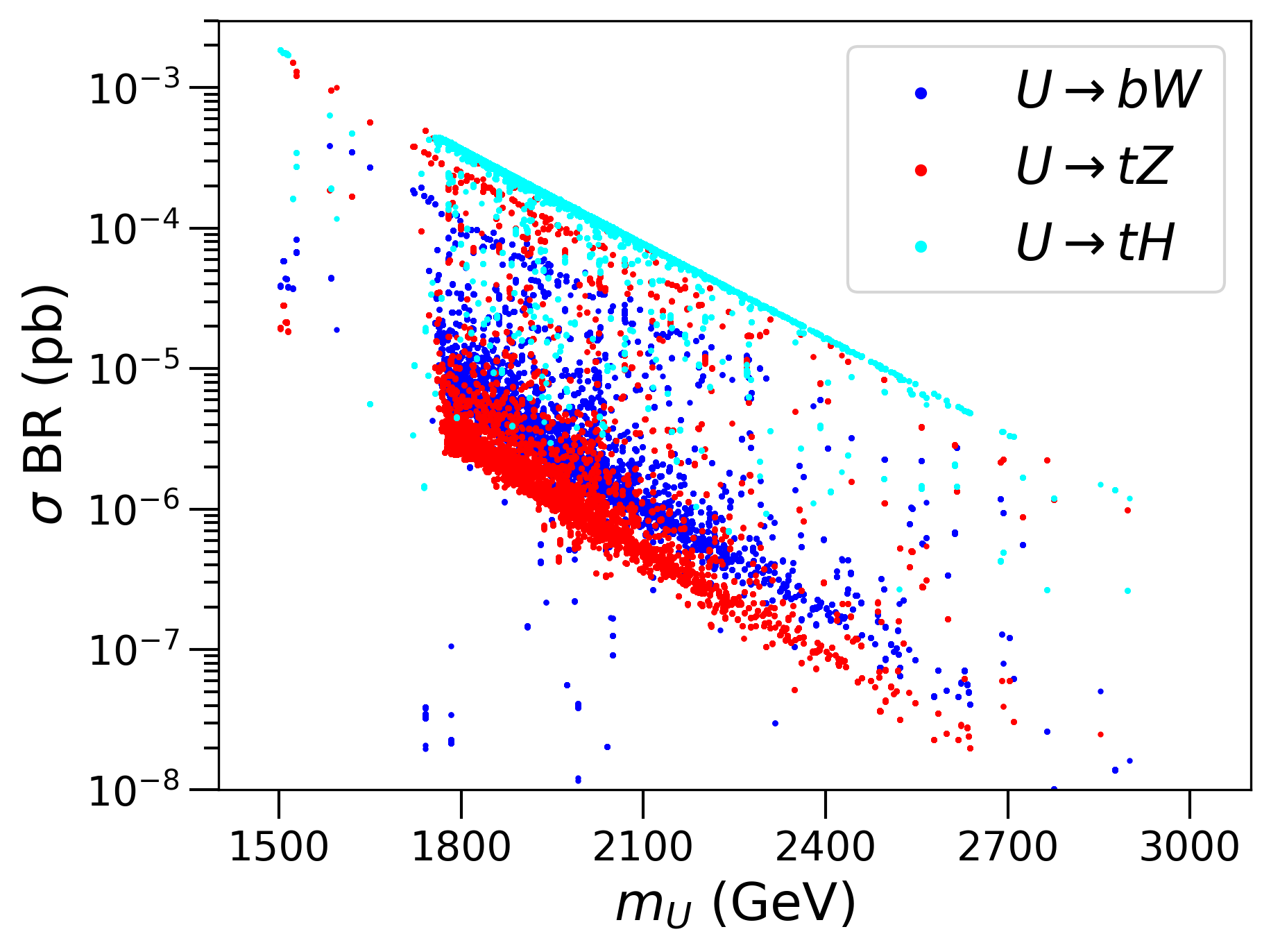}
\end{subfigure}
\begin{subfigure}{.49\linewidth}
  \centering
  \includegraphics[width=1\textwidth, clip]{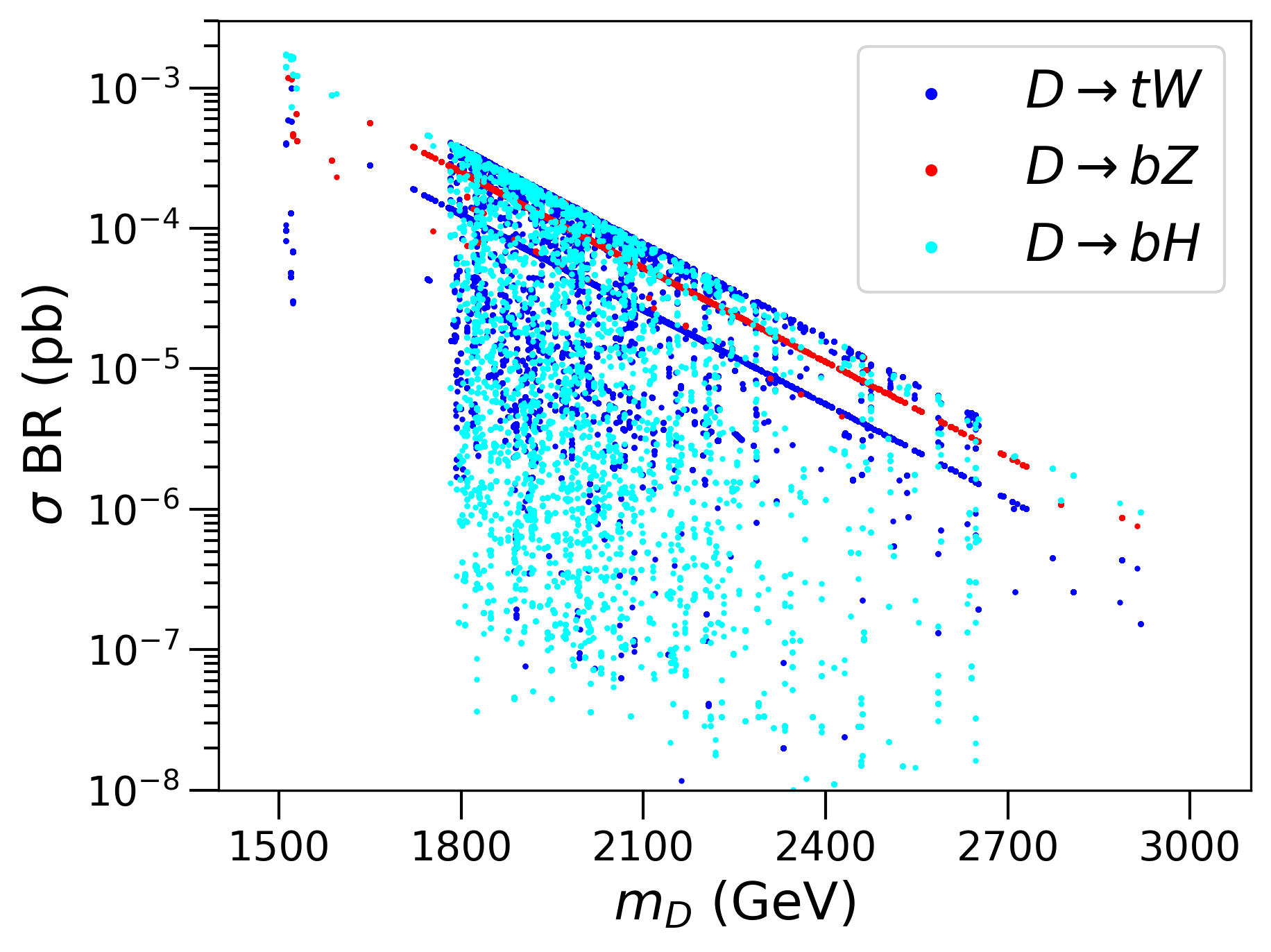}
\end{subfigure}
\begin{subfigure}{.49\linewidth}
  \centering
  \includegraphics[width=1\textwidth, clip]{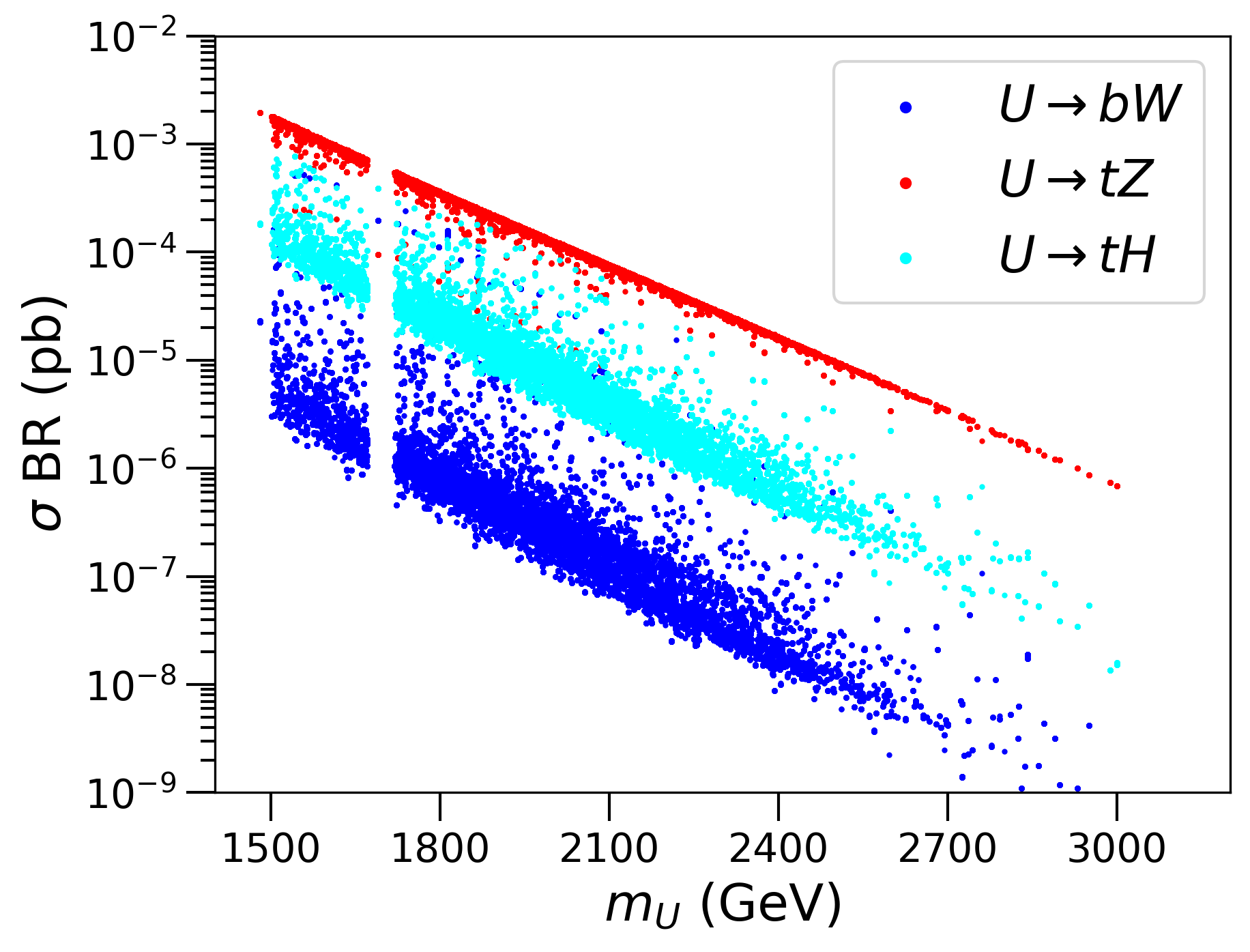}
\end{subfigure}
\begin{subfigure}{.49\linewidth}
  \centering
  \includegraphics[width=1\textwidth, clip]{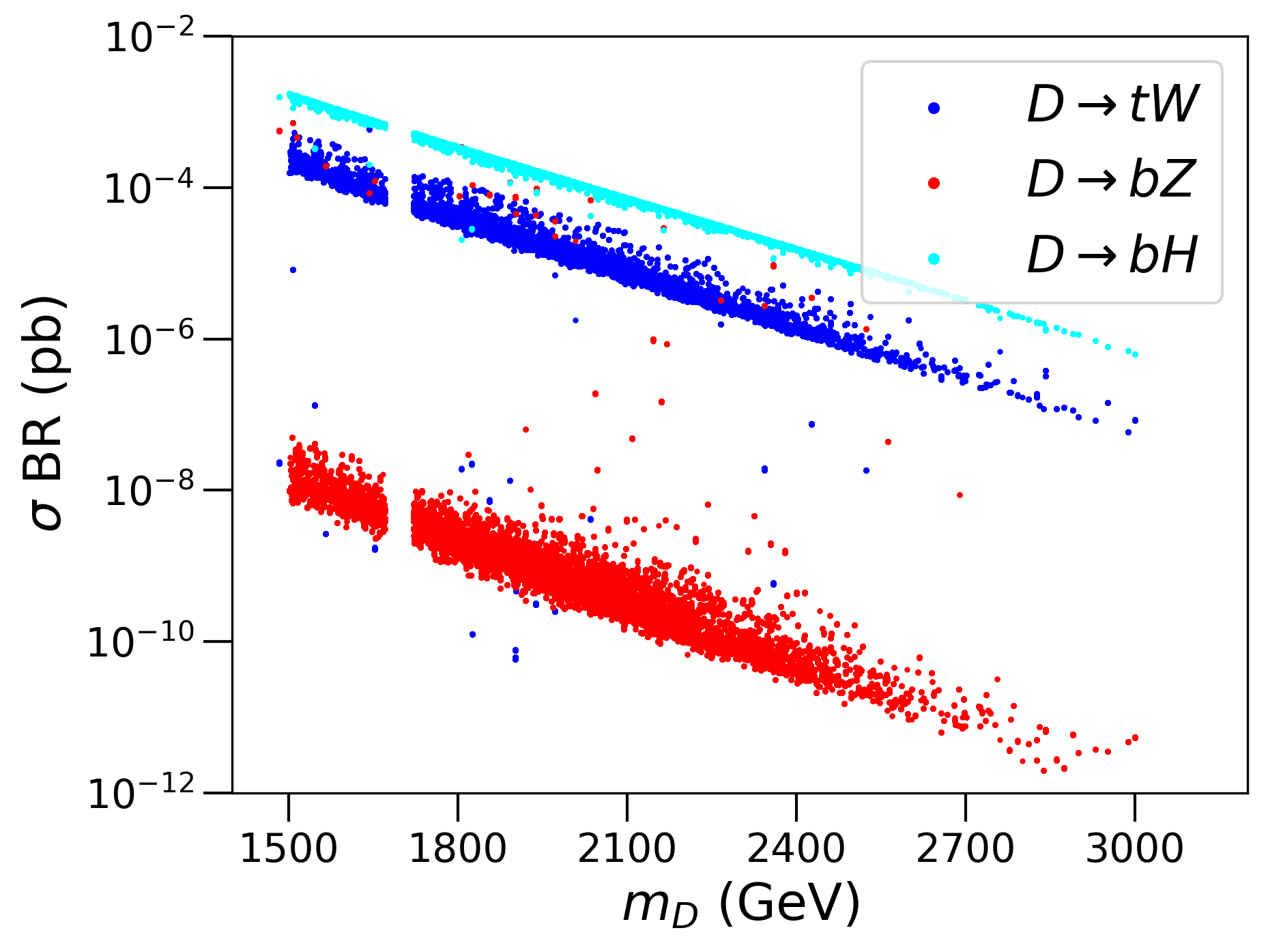}
\end{subfigure}
\caption{Cross sections for the pair-production of the lightest up-type (U) and down-type (D) resonances decaying into various final states at the $\sqrt{s}=13$~TeV LHC for viable points in (top) the M4DCHM$^{5-5-5}$, (middle) the M4DCHM$^{14-14-10}$, (bottom) the M4DCHM$^{14-1-10}$.}
\label{fig:cross_sections}
\end{figure}

Cross sections for the pair-production of these resonances, and one of the pair's subsequent decay into various Standard Model final states at the $13$~TeV LHC, are shown in Fig.~\ref{fig:cross_sections}. The points included here are all those that our scans found that satisfy each individual constraint at the $3\sigma$ level. Note these viable points lead to some resonances of lower mass than those present in Fig.~\ref{fig:mass_likelihood_profiles}, especially in the M4DCHM$^{5-5-5}$, whose double-tuning favours light composite partners. Based on Fig.~\ref{fig:cross_sections}, the U decays offer promising channels for future collider searches, having cross sections for lower-mass resonances $m_{U} \lesssim 1.6$~TeV quite near the current upper bounds, most notably in the M4DCHM$^{5-5-5}$ and to a lesser extent the M4DCHM$^{14-1-10}$. The M4DCHM$^{14-1-10}$ also offers quite clear predictions for the hierarchy of branching ratios into different final states for each species.

\section{Conclusions}
\label{Conclusions}

%
We have performed the first global fits of the M4DCHM$^{5-5-5}$, M4DCHM$^{14-14-10}$, and M4DCHM$^{14-1-10}$ minimal composite Higgs models. The M4DCHM$^{5-5-5}$ is, however, poorly sampled - possibly due to the double-tuning from which this model suffers. The former models have been found to prefer larger values of the Higgs decay constant $f$, and are therefore expected to be finely-tuned, while the latter has $f$ constrained to between roughly $1.9$~TeV and $3.2$~TeV at the $2\sigma$ confidence level. All models contain new up-type and down-type resonances with masses in ranges from approximately $1.2$~TeV to $3.0$~TeV in their viable regions, and those on the lower end of the spectrum will soon be within reach of the LHC. We plan to extend this work soon with global fits of these models in a Bayesian framework, to weigh the fitness of each region against the inherent tunings in the models.

%
\bibliography{refs}
%
%
%
%

\end{document}